\documentclass[proceedings]{rmaa}

\usepackage{paralist}

\usepackage{psfrag,color}

\newcommand{\msec}[2]{$#1\mbox{$''\mskip-7.6mu.\,$}#2$}

\SetYear{2008}
\SetConfTitle{4$^{th}$ International Meeting of Dynamical Astronomy 
in Latin America}

\title{Very Long Baseline Array astrometry \\ 
of low-mass young stellar objects} 

\author{
  Laurent Loinard,\altaffilmark{1}
  Rosa M.\ Torres,\altaffilmark{1}
  Amy J.\ Mioduszewski,\altaffilmark{2}
  and Luis F.\ Rodr\'{\i}guez\altaffilmark{1}}

\altaffiltext{1}{Centro de Radioastronom\'\i{}a y Astrof\'\i{}sica,
  UNAM, Campus Morelia, Apartado Postal 3--72, 58090, Morelia,
  Michoac\'an, M\'exico (l.loinard@astrosmo.unam.mx).}
\altaffiltext{2}{National Radio Astronomy Observatory, 
  Array Operations Center, 1003 Lopezville Road, Socorro, NM 87801, USA.}

\shortauthor{Loinard et al.}
\shorttitle{VLBA astrometry of low-mass young stellar objects}

\listofauthors{L. Loinard, R.M. Torres, A.J. Mioduszewski, \& L.F. 
Rodr\'{\i}guez}
\indexauthor{Loinard, L.}
\indexauthor{Torres, R.M.}
\indexauthor{Mioduszewski, A.J.}
\indexauthor{Rodr\'{\i}guez, L.F.}

\abstract{Multi-epoch radio-interferometric observations of young
stellar objects can be used to measure their displacement over the
celestial sphere with a level of precision that currently cannot be
attained at any other wavelength. In particular, the accuracy achieved
using carefully calibrated, phase-referenced observations with the
{\em Very Long Baseline Array} is better than 50
micro-arcseconds. This is sufficient to measure the trigonometric
parallax and the proper motion of any radio-emitting young star within
several hundred parsecs of the Sun with an accuracy better than a few
percents. Taking advantage of this situation, we have initiated a
large project aimed mainly at measuring the distance to the nearest
regions of star-formation (Taurus, Ophiuchus, Perseus, etc.). Here, we
will present the results for several stars in Taurus and Ophiuchus,
and show that the accuracy obtained is already more than one order of
magnitude better than that of previous estimates. The proper motion
obtained from the data can also provide important information,
particularly in multiple stellar systems. To illustrate this point, we
will present the case of the famous system T Tauri, where the VLBA
data provide crucial information for the characterization of the
orbital path.}

\resumen{Usando observaciones radio-interferom\'etricas obtenidas a
multiples \'epocas, se puede medir el desplazamiento de objetos
estelares j\'ovenes sobre la b\'oveda celeste con un nivel de
precisi\'on actualmente inalcanzable a cualquier otra longitud de
onda. En particular, la precisi\'on obtenida con observaciones tomadas
con el {\em Very Long Baseline Array}, usando referencia de fase,
puede ser mejor que 50 micro-segundos de arco si se sigue una
calibraci\'on cuidadosa. Eso es suficiente para medir la paralaje
trigonom\'etrica y el movimiento propio de cualquier radio-estrella
joven localizada a menos que unos cientos de parsecs del Sol con una
precisi\'on mejor que unos cuantos porcientos. Aprovechando esta
situaci\'on, hemos empezado a desarrollar un gran proyecto cuya meta
principal es medir la distancia a las regiones de formaci\'on estelar
m\'as cercanas (Tauro, Ofiuco, Cefeo, etc.). Aqu\'{\i}, presentaremos
los resultados para varias estrellas en Tauro y Ofiuco, y mostraremos
c\'omo la precisi\'on alcanzada ya es un orden de magnitud mejor que
la de estimaciones previas. Los movimientos propios obtenidos
tambi\'en son interesantes, particularmente en sistemas estelares
m\'ultiples.  Para ilustrar este punto, presentaremos el caso del
famoso sistema T Tauri, donde los datos VLBA proveen informaci\'on
crucial para la caracterizaci\'on de la \'orbita.}

\addkeyword{Astrometry}
\addkeyword{Stars: formation}
\addkeyword{Binaries: general}
\addkeyword{Radio continuum: stars}
\addkeyword{Radiation mechanisms: non-thermal}
\addkeyword{Stars: individual (T Tau, HDE~283572, HP Tau, Hubble 4, S1, 
DoAr21)}
\begin{document}
\maketitle

\section{Introduction}

Astrometric observations of young stellar objects can provide a wealth
of important information on their properties. First and foremost, an
accurate trigonometric parallax measurement is a pre-requisite to the
derivation, from observational data, of their most important
characteristics (luminosity, age, mass, etc.). Unfortunately, even in
the current post-Hipparcos era, the distance to even the nearest
star-forming regions (Taurus, Ophiuchus, Perseus, etc.) is rarely
known to better than 20 to 30\% (e.g.\ Knude \& H\"og 1998; Bertout et
al.\ 1999). At this level of accuracy, the mass of a binary system
derived from observations of its orbital motion would be uncertain by
a factor of two. This unsatisfactory state of affairs is largely the
result of the fact that young stars are still embedded in their opaque
parental cloud. They are, therefore, dim in the visible bands that
were observed by Hipparcos.

The proper motions that can be derived from astrometric observations
of young stars are also of interest. They can be used to study the
overall dynamics of star-forming regions as well as the kinematics of
the mass ejections that are often driven by young stars. If they are
scheduled to adequately characterize the orbital paths of young
multiple systems, astrometric measurements can also provide accurate
mass estimates. This is particularly important to constrain pre-main
sequence evolutionary models (e.g.\ Hillenbrand \& White 2004).

Since observations of young stars in the visible range are limited by
the effect of dust extinction, one must turn to a more favorable
wavelength regime in order to obtain high quality astrometric
data. Radio observations, particularly using large interferometers is
currently the best prospect because (i) the interstellar medium is
largely transparent at these wavelengths, and (ii) the astrometry
delivered by radio-interferometers is extremely accurate and
calibrated against fixed distant quasars. Of course, only those young
stars associated with radio sources are potential targets. Moreover,
radio interferometers effectively filter out any emission more
extended than a certain limiting angular size. For instance, in the
data presented below, only emission more compact than about 40
milli-arcseconds can be detected. In young stars, only non-thermal
emission mechanisms can generate detectable emission on such compact
scales, so we must concentrate on magnetically active sources. This is
not a particularly limiting factor, fortunately, because low-mass
young stars do tend to have intense surface magnetic fields (e.g.\
Johns-Krull 2007) that can generate detectable non-thermal emission.

\section{Observations}

In this paper, we will consider a total of six young stars: four in
Taurus (T Tauri Sb, HDE~283572, Hubble 4 and HP Tau/G2) and two in
Ophiuchus (DoAr 21 and S1). All six objects were previously known
non-thermal radio sources. Each was observed at 6 to 12 epochs
separated from one another by 2 to 3 months, and covering a total
timespan of 1.25 to 2.5 years. All the observations were collected at
a wavelength of 3.6 cm with the {\em Very Long Baseline Array}, an
interferometer of 10 antennas, each 25 meters in diameter, spread over
the entire US territory (see http://www.vlba.nrao.edu/ for
details). Phase-referencing --whereby observations of the scientific
target and a nearby quasar used as calibrator are intertwined-- was
used for all the data. The calibration scheme is described in detail
in Loinard et al.\ (2007a) and Torres et al.\ (2007).

The typical angular resolution of our data is 1 milli-arcsecond, and
the typical astrometric accuracy is 50 to 100 micro-arcseconds.  The
final noise level in the images is 50 to 100 mJy, sufficient to
always detect the sources at more than about 10$\sigma$. The source
flux density varies from object to object and epoch to epoch, from a
minimum of about 0.5 mJy to a maximum of nearly 50 mJy.

\begin{figure*}[!t]
\centerline{\includegraphics[width=1.6\columnwidth,angle=270]{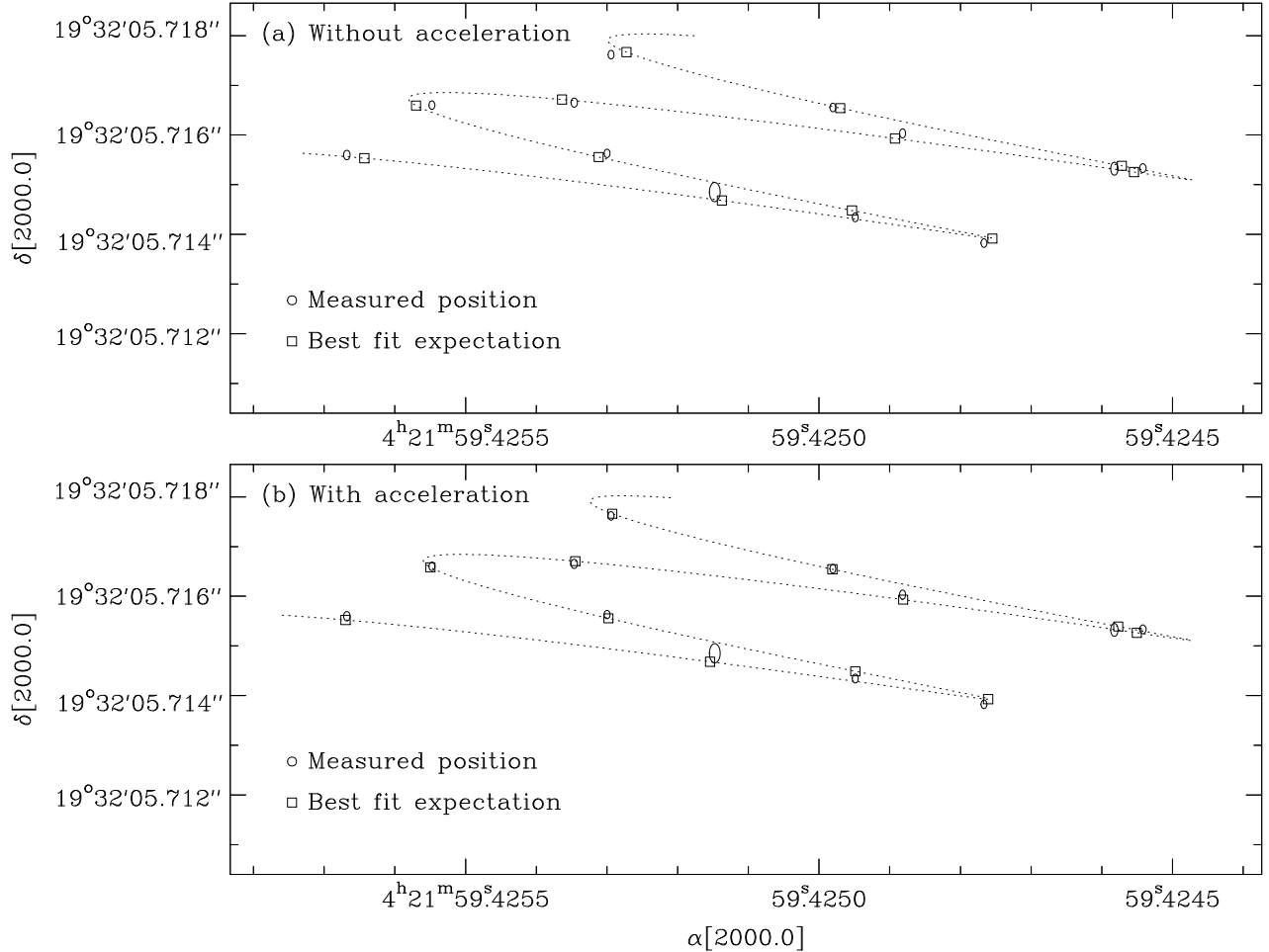}}
  \caption{Measured positions of T Tau Sb and best fit without (a) and
  with (b) acceleration terms. The observed positions are shown as
  ellipses, the size of which represents the magnitude of the
  errors. Note the very significant improvement when acceleration
  terms are included.}  
\end{figure*}

\section{Distance to the nearest star-forming regions}

The displacement of a source on the celestial sphere is the
combination of its trigonometric parallax ($\pi$) and its proper
motion. In what follows, we will have to consider three different
situations in terms of proper motions. Two of our target stars are
apparently single (Hubble 4 and HDE~283572), and one (HP~Tau/G2) is a
member of a multiple system with an orbital period so much longer than
the timespan covered by the observations that the effect of the
companions can safely be ignored. Thus, in these three cases, the
proper motion can be assumed to be linear and uniform, and the right
ascension ($\alpha$) and the declination ($\delta$) vary as a function
of time $t$ as:

\begin{eqnarray}
\alpha(t) & = & \alpha_0+(\mu_\alpha \cos \delta) t + \pi f_\alpha(t) \label{uni1}\\%
\delta(t) & = & \delta_0+\mu_\delta t + \pi f_\delta(t), \label{uni2}
\end{eqnarray}

\noindent where $\alpha_0$ and $\delta_0$ are the coordinates of the
source at a given reference epoch, $\mu_\alpha$ and $\mu_\delta$ are
the components of the proper motion, and $f_\alpha$ and $f_\delta$ are
the projections over $\alpha$ and $\delta$, respectively, of the
parallactic ellipse. 

On the other hand, two of our sources (DoAr~21 and S1) are members of
compact binary systems with an orbital period of the order of the
timespan covered by the observations. In such a situation, one should
fit simultaneously for the uniform proper motion of the center of mass
and for the Keplerian orbit of the system. This, however, requires
more observations than are needed to fit only for a uniform proper
motion. Thus, additional data are currently being collected to
adequately constrain the required fits. In this paper, we will present
preliminary results based on Eqs.\ (1) and (2) where the Keplerian
motion is not included. We will see momentarily that the main effect
of not fitting for the Keplerian orbit is an increase in the final
uncertainty on the distance.

Finally, one source (T Tauri Sb) is a member of a binary system with
an orbital period longer than the timespan of our observations but not
by a huge factor. While a full Keplerian fit would again, in
principle, be needed, we found that including a constant acceleration
term provides an adequate description of the trajectory. The fitting
functions in this case are of the form:

{\footnotesize

\begin{eqnarray}
\alpha(t) & = & \alpha_0+(\mu_{\alpha 0} \cos \delta) t + {1 \over 2} (a_\alpha \cos \delta) t^2 + \pi f_\alpha(t) \label{acc1}\\%
\delta(t) & = & \delta_0+\mu_{\delta 0} t + {1 \over 2} a_\delta t^2 + \pi f_\delta(t), \label{acc2}
\end{eqnarray}

}

\noindent where $\mu_{\alpha 0}$ and $\mu_{\delta 0}$ are the proper
motions at a reference epoch, and $a_\alpha$ and $a_\delta$ are the
projections of the uniform acceleration.

\begin{figure*}[!t]
\centerline{\includegraphics[width=1.3\columnwidth,angle=270]{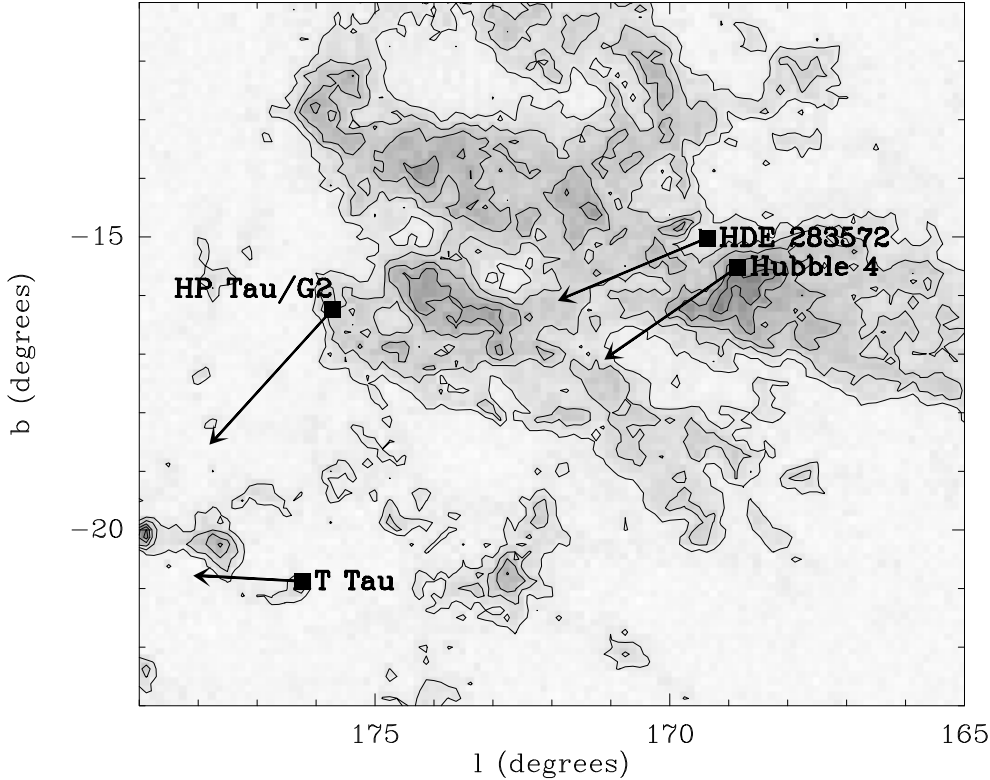}}
  \caption{Position and tangential velocity vectors of the four young
  stars in Taurus considered here, superimposed onto the CO(1-0) map of
  Taurus from Dame et al.\ (2001).}  
\end{figure*}

\subsection{Distance to Taurus}

In total, twelve observations of T Tau Sb were obtained between
September 2003 and July 2005. The trajectory described by the source
on the plane of the sky (Fig.\ 1) can be adequately described as the
superposition of a parallactic ellipse and a uniformly accelerated
proper motion. The resulting parallax is 6.82 $\pm$ 0.03 mas,
corresponding to a distance of 146.7 $\pm$ 0.6 pc (Loinard et al.\
2005, 2007a; Tab.\ 1). For Hubble 4 and HDE~283572, six observations
were obtained between September 2004 and December 2005. Fits with a
uniform proper motion adequately described the sources trajectories
(Torres et al.\ 2007), and the resulting parallaxes are 7.53 $\pm$
0.03 mas ($d$ = 132.8 $\pm$ 0.5 pc) and 7.78 $\pm$ 0.04 mas ($d$ =
128.5 $\pm$ 0.6 pc), respectively (Tab.\ 1). Finally, nine
observations of HP~Tau/G2 were obtained between September 2005 and
December 2007. Again, a uniform proper motion was sufficient to
properly describe the source path. The best fit yields a parallax of
6.20 $\pm$ 0.03 mas, corresponding to 161.2 $\pm$ 0.9 pc (Torres et
al.\ 2008, in prep.; Tab.\ 1).

Two of these four objects have measured Hipparcos parallaxes (Bertout
et al.\ 1999): T Tau ($\pi_{hip} = 5.66 \pm 1.58$) and HDE~283572
($\pi_{hip} = 7.81 \pm 1.30$). Our results are consistent with these
values, but one to two orders of magnitude more accurate. Also, the
trigonometric parallax to both Hubble 4 and HDE~283572 was estimated
by Bertout \& Genova (2006) using a modified convergent point
method. Their results ($\pi_{cp} = 8.12 \pm 1.50$ for Hubble 4 and
$\pi_{cp} = 7.64 \pm 1.05$ for HDE~283572) are also consistent with
our results, but again more than one order of magnitude less accurate.
Indeed, our measurements yield distances accurate to about 0.5\% for
all four sources.

Taking the weighted mean of our four measurements, we can estimate the
mean parallax of the Taurus association to be $\bar{\pi}$ = 7.0 mas,
corresponding to a mean distance $\bar{d}$ = 143 pc. This is in good
agreement with the value of 140 $\pm$ 15 pc traditionally used for
Taurus (e.g.\ Kenyon et al.\ 1994, Bertout et al.\ 1999).

As can be seen from Fig.\ 2, the total spatial extent of Taurus on the
sky is about 10$^\circ$, corresponding to a physical size of about 25
pc. Our observations show that the depth of the complex is similar
since HP Tau is about 30 pc farther than Hubble 4 or
HDE~283572. Because of this significant depth, even if the mean
distance of the Taurus association were known to infinite accuracy, we
could still make errors as large as 10--20\% by using the mean
distance indiscriminately for all sources in Taurus. To reduce this
systematic source of error, one needs to establish the
three-dimensional structure of the Taurus association, and
observations similar to those presented here currently represent the
most promising avenue toward that goal. Indeed, the observations of
the four stars presented earlier already provide some hints of what
the three-dimensional structure of Taurus might be. Hubble 4 and
HDE~283572 which were found to both be at about 130 pc and to share a
similar kinematics (Torres et al.\ 2007 --see Fig.\ 2), are also
located in the same portion of Taurus, near Lynds 1495. T Tauri is
located in the southern part of Taurus near Lynds 1551, its tangential
velocity is clearly different from that of Hubble 4 and HDE~283572,
and it appears to be somewhat farther from us. Finally, HP Tau is
located near the (Galactic) eastern edge of Taurus, and is the
farthest of the four sources considered here. Although additional
observations are needed to draw definite conclusions, our data,
therefore, suggest that the region around Lynds 1495 corresponds to
the near side of the Taurus complex at about 130 pc, while the eastern
side of Taurus corresponds to the far side at 160 pc. The region
around Lynds 1551 and T Tauri appears to be at an intermediate
distance of about 147 pc.

\begin{figure*}[!t]
\centerline{\includegraphics[width=1.0\columnwidth,angle=270]{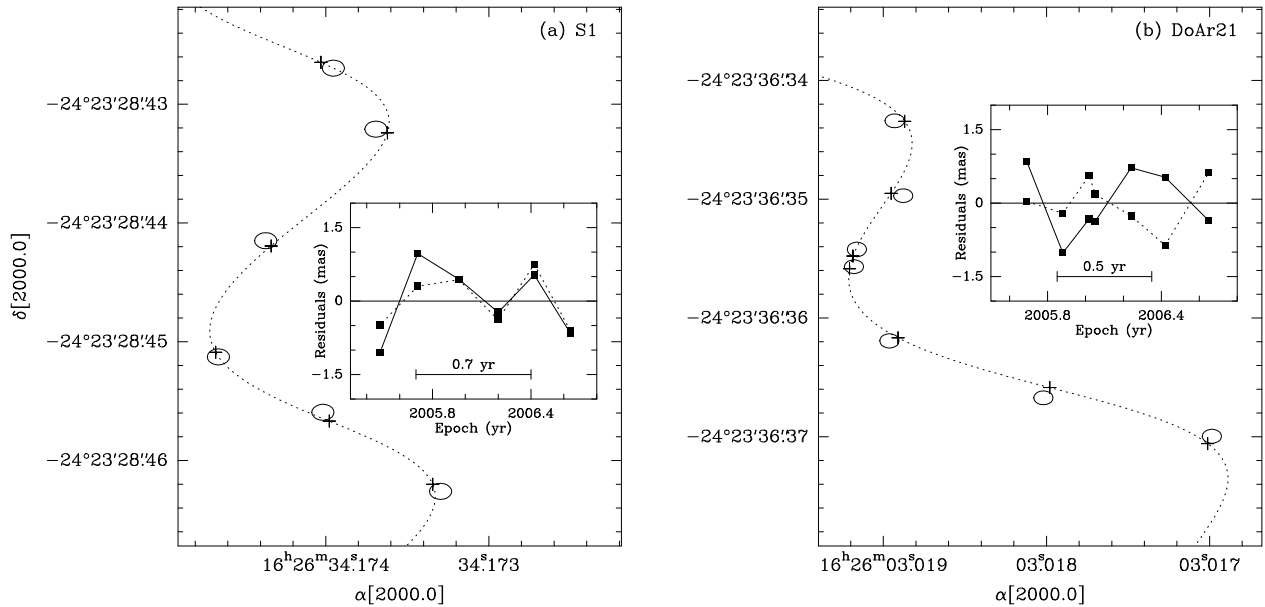}}
  \caption{Measured positions and best fit for (a) S1, and (b) DoAr21.
  The observed positions are shown as ellipses, the size of which
  represents the magnitude of the errors. The positions at each epoch
  expected from the best fits are shown as $+$ signs. The insets show
  the residuals (fit--observation) in right ascension (full line) and
  declination (dotted line).}  
\end{figure*}

\begin{table}[!t]
\caption{Parallaxes and Distances to the six stars considered here}
\begin{tabular}{lcc}
Source & parallax (mas) & Distance (pc) \\%
\hline
Taurus: \\%
T Tau Sb   & 6.82 $\pm$ 0.03 & 146.7 $\pm$ 0.6 \\%
Hubble 4   & 7.53 $\pm$ 0.03 & 132.8 $\pm$ 0.5 \\%
HDE~283572 & 7.78 $\pm$ 0.04 & 128.5 $\pm$ 0.6 \\%
HP Tau     & 6.20 $\pm$ 0.03 & 161.2 $\pm$ 0.9 \\%
Ophiuchus: \\%
S1         & 8.55 $\pm$ 0.50 & 116.9$^{+7.2}_{-6.4}$ \\%
DoAr21     & 8.20 $\pm$ 0.37 & 121.9$^{+5.8}_{-5.3}$ \\%
\hline
\end{tabular}
\end{table}

\subsection{Distance to the Ophiuchus core}

Ophiuchus is, together with Taurus, one the best studied regions of
low-mass star-formation. Unlike Taurus, it is only a few parsecs
across so a reliable distance determination to a few of its stars
would automatically provide an accurate distance to the entire region.
Ophiuchus has long been thought to be at 165 pc (Chini 1981), but
somewhat shorter distances (120--135 pc) have recently been proposed
(Knude \& H\"og 1998; Mamajek et al.\ 2007).

Six observations of S1 and 7 observations of DoAr 21 were obtained
between June 2005 and August 2006 (Loinard et al.\ 2008). Their
observed trajectories can be adequately described as a combination of
a parallactic ellipse and a uniform proper motion, but with a post-fit
r.m.s.\ significantly larger than in the case of Taurus (Fig.\
3). Moreover, the residuals do not appear to be random, but instead
show some indication of periodicity (see the insets of Fig.\ 3), as
would be expected if the studied sources were members of binary
systems. Interestingly, S1 was previously known to have a nearby
companion (Richichi et al.\ 1994), and the expected orbital period of
the system is comparable to the period observed in Fig.\ 3 (see
Loinard et al.\ 2008 for details). DoAr21 was not previously known to
be a multiple stellar system, but was found to be a double radio
source in one of our observations. We conclude that the fairly large
periodic residuals observed for S1 and DoAr 21 are a consequence of
their unmodeled Keplerian motion. Additional VLBA observations are
being collected to properly constrain the orbital paths.

The parallaxes that are obtained from the fits shown in Fig.\ 3 are
8.55 $\pm$ 0.50 mas and 8.20 $\pm$ 0.37 mas for S1 and DoAr 21,
respectively. The corresponding distances are 116.9$^{+7.2}_{-6.4}$ pc
and 121.9$^{+5.8}_{-5.3}$ pc. Thus, the uncertainties on the distances
to the stars in Ophiuchus are significantly larger than those for the
stars in Taurus. This is a consequence of the larger post-fit r.m.s.\
mentioned earlier, and ought to be drastically reduced once full
Keplerian fits are performed. The weighted mean of the two parallaxes
is 8.33 $\pm$ 0.30, corresponding to a distance of
120.0$^{+4.5}_{-4.3}$ pc. This is in good agreement with the value
proposed by Knude \& H\"og (1998) and, more recently, by Lombardi et
al.\ (2008). Note that, in spite of the enhanced errors related to the
binarity of the sources, our determination of the distance to
Ophiuchus is accurate to better than 4\% ($\equiv$ 5 pc). This is to
be compared to the situation prior to our observations, when the
uncertainty was between 120, 140 or 160 pc.

\section{On the orbital motion of T Tau Sb}

The eponym of an entire class of young stellar objects, T Tauri was
first noticed for its variability by the British astronomer John
Russell Hind in 1852. It was first recognized to be a young star by
Joy (1945), and has been the subject of numerous studies ever since.
Initially thought to be a single star, it was discovered by Dyck et
al.\ (1982) to have a companion located about \msec{0}{7} to its
south. As a consequence, the {\em historical} T Tauri star is now
usually referred to as T Tau N, and the southern companion as T Tau S.
Recently, T Tau S was itself discovered to be a tight binary (Koresko
2000; K\"ohler et al.\ 2000) formed by two stars called T Tau Sa and T
Tau Sb, in fairly rapid relative motion (Duch\^ene et al.\ 2002, 2005,
2006; Schaefer et al.\ 2006; K\"ohler et al.\ 2008).

The structure of T Tau at radio wavelengths is somewhat complex
(Loinard et al.\ 2007b). T Tau N is associated with a compact radio
source tracing the base of its jet (e.g.\ Johnston et al.\ 2003). The
southern companion, on the other hand, is comprised of a compact
non-thermal source associated with the active magnetosphere of T Tau
Sb, and an extended thermal halo presumably related to stellar winds
(Loinard et al.\ 2007b). 

The nature of the orbital motion between T Tau Sa and T Tau Sb has
been somewhat disputed in recent years. Using 20 years worth of VLA
observations, Loinard et al.\ (2003) concluded that the orbit between
T Tau Sa and T Tau Sb had been dramatically altered after a recent
perioastron passage. More recently, however, Duch\^ene et al.\ (2006)
argued that these VLA observations, together with newer infrared data,
could be adequately described by a single Keplerian orbit with a period
of about 22 years. Finally, K\"ohler et al.\ (2008) found that the
infrared data alone were best described by a Keplerian orbit with a
period of about 90 years, and that the most recent infrared data were
not easily reproduced by the fit proposed by Duch\^ene et al.\ (2006).
The orbit favored by K\"ohler et al.\ (2008), however, is unable to
explain the older radio observations.

\begin{figure*}[!t]
\centerline{\includegraphics[width=1.35\columnwidth,angle=270]{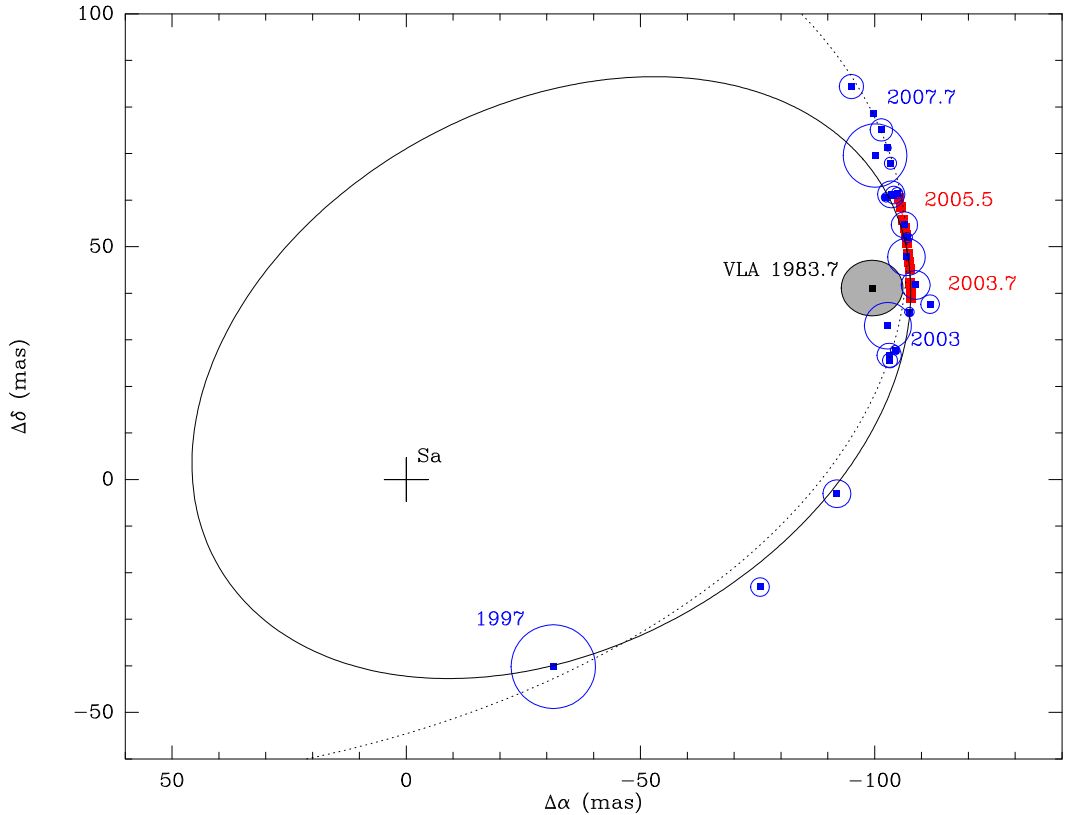}}
  \caption{Orbit of the T Tau Sa/T Tau Sb system. The blue symbols
  correspond to all the infrared observations available at the
  beginning of 2008. All the data tabulated by K\"ohler et al.\ (2008)
  as well as more
  recent Keck observations kindly provided by G.\ Schaefer are included. 
The red
  dots show are the VLBA observations. The full line correspond to the
  fit published by Duch\^ene et al.\ (2006) and the dotted line to the
  fit proposed by K\"ohler et al.\ (2008). The position of the VLA
  source in 1983.7 is shown as a grey symbol.}  \end{figure*}

Since T Tau Sb is a member of a triple stellar system, its trajectory
on the plane of the sky results from the combination of several terms:
its trigonometric parallax; the overall proper motion of the T Tauri
system relative to the Sun; the Keplerian orbit of T Tau S about T Tau
N; and the relative motion between T Tau Sb and T Tau Sa. The
trigonometric parallax has been accurately measured using the VLBA
observations (see Sect.\ 3), and can easily be removed. The overall
proper motion of the T Tauri system relative to the Sun as well as the
orbit of T Tau S about T Tau N are well determined, and can also
be taken into account (see Loinard et al.\ 2007a for
details). Thus, our VLBA observations of T Tau Sb allow us to trace
the relative motion between T Tau Sb and T Tau Sa for the period
2003.7 to 2005.5. The agreement between our radio observations and the
infrared data for the same period is excellent (Fig.\ 4), and the
radio data can be used to further constrain the fits to the orbital
path of the system. Indeed, the scatter in the radio data is less than
that in the infrared observations, so the radio position provide very
strong constraints.

Also shown in Fig.\ 4 are the fits proposed by Duch\^ene et al.\
(2006) and K\"ohler et al.\ (2008). As noticed already by Loinard et
al.\ (2007a), the VLBA data favor an orbital period somewhat longer
than that proposed by Duch\^ene et al.\ (2006), and in better
agreement with the most recent infrared observations and with the fit
proposed by K\"ohler et al.\ (2008). The difficulty with such a long
orbital period is related to the older VLA positions of T Tau S
reported by Loinard et al.\ (2003). To illustrate this problem, we
show in Fig.\ 4 the position of the VLA source associated with T Tau S
around 1984. Clearly, the observed VLA position would be consistent
with the 20--22 yr orbital period proposed by Duch\^ene et al.\
(2006): between the old VLA observation and the IR or VLBA data
corresponding to 2003--2005, T Tau Sb would have complete a full
orbit. For a 90 yr orbital period, however, the position expected for
1984 is located at about $\Delta \alpha$ $\sim$ +100 mas; $\Delta
\delta$ $\sim$ --70 mas. This is clearly inconsistent with the VLA
position observed at that epoch. Additional observations (infrared,
VLA, and VLBA) are clearly needed to settle this issue. 

\section{Conclusions and perspectives}

In this paper, we presented the results of multi-epoch VLBA
observations of several low-mass young stars, which allowed us to
measure their trigonometric parallaxes with unprecedented accuracy,
and --in some cases-- study their orbital motions. Our main
conclusions, and some perspectives are presented in the following.

\subsection{Distance and structure of Taurus}

Using observations of four stars in Taurus, we confirmed that the mean
distance to this important region of star-formation is about 143
pc. In addition, the total depth of Taurus could be estimated to be
about 30 pc, and the first indications of the three-dimensional
structure of the complex could be obtained. Exploring further the
spatial structure of Taurus is very important because, given its
depth, using a mean distance indiscriminately for all Taurus members
could result in systematic errors of 10--20\%. Observations similar to
those presented here of a larger sample of stars, currently represent
the best avenue toward an accurate determination of the
three-dimensional structure of Taurus. Nowadays, however, only a few
stars in Taurus --besides those considered here-- are known
non-thermal radio emitters.  Observations of these few additional
cases are already underway or will be obtained soon. To increase
further the sample of possible candidates for multi-epoch VLBA
observations, identification of new magnetically active objects with
adequate non-thermal emission will be required.

\subsection{The distance to Ophiuchus}

The mean distance to the Ophiuchus core was also measured using
multi-epoch VLBA observations, and was found to be 120$^{+4.5}_{-4.3}$
pc. This value is in good agreement with several recent determinations
(Knude \& H\"og 1998; Lombardi et al.\ 2008) and represents a very
significant advance, since the distance to Ophiuchus was previously
uncertain by about 20 pc. Our distance determination ought to be
further improved once additional data designed to characterize the
orbital motion of the sources are available. These data are currently
being obtained and reduced. Since the total extent of the Ophiuchus
core on the plane of the sky is only a few parsecs, there is little
need for a larger sample of stars. However, several large extensions
(known as {\em streamers}) are known to exist around the Ophiuchus
core, and those could be at somewhat different distances. Since these
streamers harbor a number of interesting young stellar sources, it
would be interesting to also measure their distances accurately.

\subsection{Other regions}

Several other nearby star-forming regions have been studied in detail
at many wavelengths but have poorly determined distances (e.g.\
Perseus, Serpens, etc.). Non-thermal sources are known to exist in
these regions, so multi-epoch VLBA observations would allow
significant improvements in the determination of their distances. The
corresponding observations are being obtained and will be published in
forthcoming papers.

Note that the distance to the nearest region of high-mass
star-formation (Orion) has recently been measured using Very Long
Baseline Interferometry observations similar to those presented here 
(Menten et al.\ 2007; Hirota et al.\ 2007; Sandstrom et al.\ 2007).

\subsection{Orbital motions and mass determinations}

Finally, our observations also allow us to measure the proper motions
of the sources under study. This is particularly interesting to study
the orbital motion of multiple stellar systems. In the case of T
Tauri, we showed that the VLBA data are in excellent agreement with
infrared observations taken over the same time period, and that they
already provide very important constraints for the orbital
fits. Future observations will allow further improvements in the
determination of the orbital elements, and --combined with archival
VLA data-- will help resolve the existing dispute on the nature of the
orbital path. Eventually, the mass of the individual stars will be
measured very accurately; this will provide very important constraints
for pre-main sequence evolutionary models.

Another very interesting case is that of V773 Tau which is described
in Torres et al.\ (this volume). There, the VLBA observations
spatially resolve the two members of a tight spectroscopic binary with
a period of only about 50 days. The combination of spectroscopic
observations at visible wavelengths and multi-epoch VLBA astrometric
data will allow us to fully characterize the orbital elements and the
mass of the stars.

\acknowledgements{L.L., R.M.T, and L.F.R. acknowledge the financial
support of DGAPA, UNAM and CONACyT, M\'exico. NRAO is a facility of
the National Science Foundation operated under cooperative agreement
by Associated Universities, Inc. We are grateful to Gail Shaefer for
providing us with recent Keck observations of the T Tau System prior
to publication.}

\end{document}